\documentclass[fleqn,usenatbib]{mnras_tex_edited}

\usepackage{amssymb,amsmath,verbatim,mathtools,needspace,enumitem,etoolbox,graphicx,physics,microtype,natbib,url,hyperref,mathrsfs,bm,ulem,orcidlink}

\renewcommand{\emph}[1]{\textit{#1}}
\newcommand{\approptoinn}[2]{\mathrel{\vcenter{
  \offinterlineskip\halign{\hfil$##$\cr
    #1\propto\cr\noalign{\kern2pt}#1\sim\cr\noalign{\kern-2pt}}}}}

\DeclareRobustCommand{\okina}{%
  \raisebox{\dimexpr\fontcharht\font`A-\height}{%
    \scalebox{0.8}{`}%
  }%
}

\newcommand{\msun}{{\rm M}_\odot}

\title[The upper edge of the black hole mass gap]{The location of the upper edge of the pair-instability supernovae black hole mass gap}

\author[Sakstein \& Croon]{%
Jeremy Sakstein$\,$\orcidlink{0000-0002-9780-0922}$^{1}$
\thanks{sakstein@hawaii.edu}
Djuna Croon$\,$\orcidlink{0000-0003-3359-3706}$^{2}$
\thanks{djuna.l.croon@durham.ac.uk}
\medskip
\\
$^{1}$ Department of Physics \& Astronomy, University of Hawai\okina i, Watanabe Hall, 2505 Correa Road, Honolulu, HI, 96822, USA\\
$^{2}$ Institute for Particle Physics Phenomenology, Department of Physics, Durham University, Durham DH1 3LE, UK
}

\newcommand{\preprintnumber}{IPPP/26/54}
\pubyear{\the\year{}}

\begin{document}

\label{firstpage}
\pagerange{\pageref{firstpage}--\pageref{lastpage}}

\maketitle

\begin{abstract}
Gravitational wave observations are beginning to probe the upper edge of the pair-instability supernova (PISN) black hole mass gap, a key prediction of stellar evolution.~In this work, we quantify the sensitivity of this boundary to uncertainties in stellar evolution using a suite of simulations that vary inputs including nuclear reaction rates, mixing processes, and stellar winds.~We find that the $^{12}{\rm C}(\alpha,\gamma)^{16}{\rm O}$ reaction rate is the dominant source of uncertainty, shifting the upper edge by $\Delta M\sim30\,\msun$, with the triple-$\alpha$ rate producing a comparable shift of $\sim25\,\msun$.~Notably, $^{16}{\rm O}+^{16}{\rm O}$ reactions shift the upper edge by $\sim15\,\msun$ while leaving the lower edge unchanged, implying they can widen or narrow the mass gap.~Other processes affect the location at the $\lesssim10\,\msun$ level.~In contrast to the lower edge, we find that the upper edge is robust to variations in spatial and temporal resolution, indicating that it is reliably resolved in current simulations.~Our results demonstrate that the upper edge carries substantial theoretical uncertainty and, while comparatively less affected by astrophysical contamination than the lower edge, provides a direct probe of the nuclear processes governing pair instability.~We discuss the implications for interpreting high-mass black hole detections in gravitational wave data.
\end{abstract}

\begin{keywords}
black hole mergers --- gravitational waves
{\qquad\qquad\qquad\qquad\qquad   \preprintnumber}
\end{keywords}

\section{Introduction}

\begin{figure}
    \centering
    \includegraphics[width=0.95\linewidth]{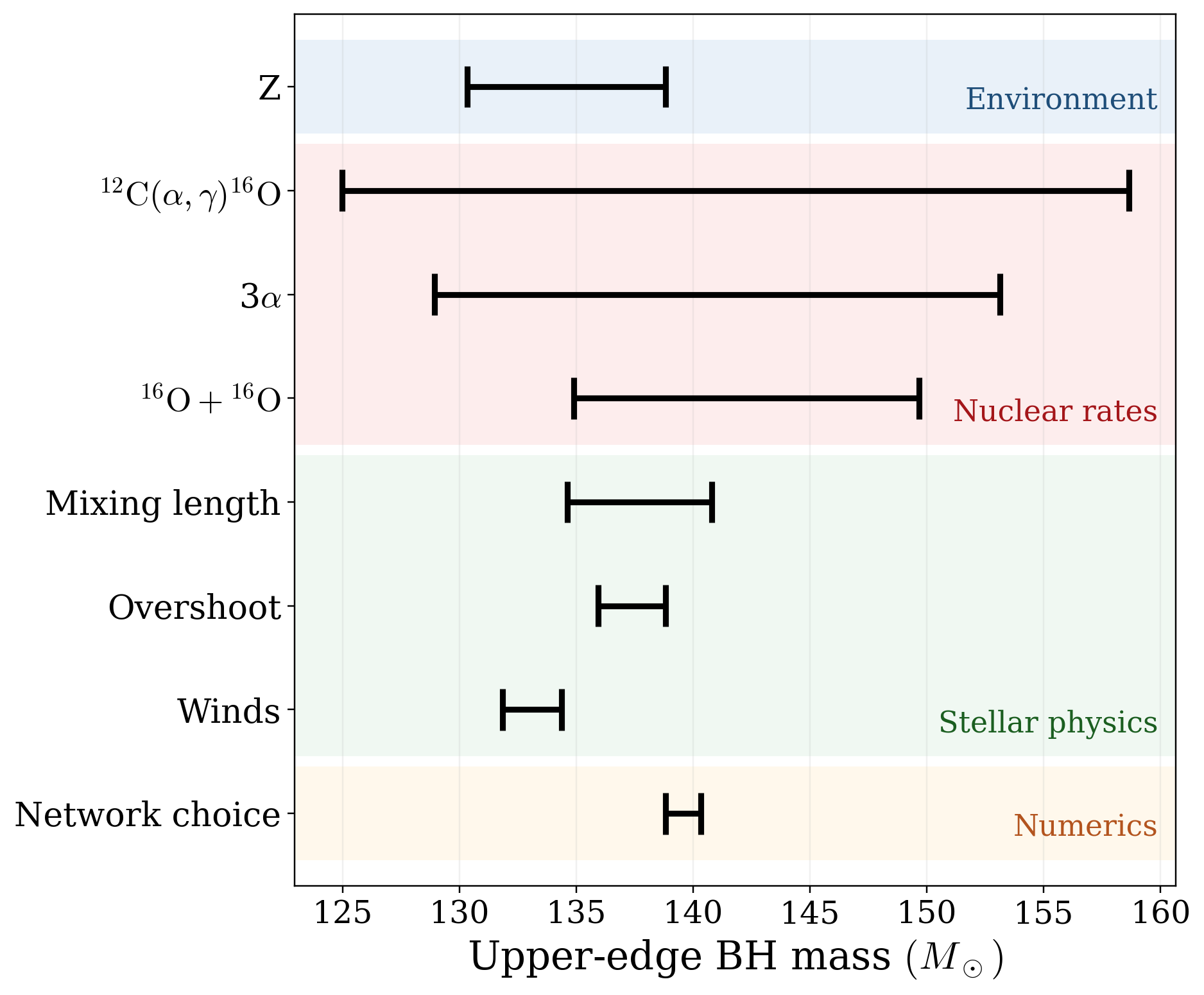}
    \caption{Largest shifts in the upper-edge BH mass under variations of individual physical inputs and modeling assumptions around a fiducial configuration with $Z=10^{-5}$ ($Z=5 \times 10^{-4}$ for winds).~Further details in text.
    }\vspace{-5mm}
    \label{fig:biggestshifts}
\end{figure}

Stellar evolution theory predicts a black hole mass gap (BHMG) --- a range of black hole (BH) masses that cannot be formed through the direct core collapse of massive stars, due to their total disruption by pair-instability supernovae (PISNe) \citep{Barkat:1967zz,1968Ap&SS...2...96F,1967ApJ...148..803R,2002RvMP...74.1015W,Belczynski:2016jno,2017ApJ...836..244W,2019ApJ...878...49W}.~The location of the lower edge is uncertain, depending on stellar physics such as nuclear reaction rates \citep{2019ApJ...887...53F,Farmer:2020xne,Mehta:2021fgz} and on numerical resolution \citep{Farag:2022jcc}; however, stellar structure simulations generally predict it to lie in the range $45$--$90\,\msun$, placing it within the reach of current gravitational wave (GW) interferometers.~The upper edge is expected to occur at $M\gtrsim120\,\msun$ \citep{Marchant:2018kun,Farmer:2020xne,Mehta:2021fgz,Farag:2022jcc,Xin:2026zyt} and is also coming within observational reach.~To date, two possible upper mass gap objects have been detected \citep{Fishbach:2020qag,Croon:2025gol}.~Motivated by the prospect of detecting upper-edge objects in the near future, in this work we undertake a systematic numerical study of the location of the upper edge, focusing on its sensitivity to nuclear and stellar physics and numerical resolution.

Our findings are summarized in Fig.~\ref{fig:biggestshifts}, which shows the range of inferred upper-edge BH masses obtained when varying each process independently from a fiducial model.~The dominant uncertainty arises from nuclear reaction rates, with variations in $^{12}\mathrm{C}(\alpha,\gamma)^{16}\mathrm{O}$ and the $3\alpha$ process producing the largest shifts.~In contrast, metallicity $Z$ and other stellar physics inputs introduce subdominant changes.~Parameters whose variations shift the upper edge by {$<1\,\msun$ are not shown}.~These results indicate that, as GW observations begin to access this BH mass regime, the upper edge of the BHMG can provide a complementary probe of stellar microphysics, in particular the uncertain nuclear reaction rates that govern advanced burning stages.~

\section{Origin of the upper mass gap edge}

The BHMG originates from pair-instability in the late stages of stellar evolution, and subsequent explosive burning.~Post-core helium burning, massive stars ($M\gtrsim20\,\msun$) contract, raising their central temperatures and densities to values where electron-positron pairs can be thermally produced efficiently in stellar cores.~These are produced non-relativistically, such that they contribute negligibly to the core's radiative pressure support.~Consequently, the first adiabatic index falls below the critical value of $4/3$, initiating a collapse.~The subsequent increase in temperature causes explosive core oxygen ignition, releasing enough energy to unbind the entire star, leaving no remnant.~Lower mass objects can be partially disrupted, resulting in extreme mass loss episodes.~Stars undergoing such pulsational pair-instability supernovae leave remnants with significantly reduced mass.~Sufficiently light stars evade the instability and form black holes with masses close to their carbon-oxygen core mass.~The \emph{lower edge} of the BHMG corresponds to the highest mass BH that can be formed from the interplay of these processes.

{Pair-instability is expected to operate in all stellar objects above a  critical mass, but for sufficiently massive stars the resulting thermonuclear burning is unable to unbind the star.}
For these stars, the contraction heats the core to temperatures at which the photodisintegration of heavy nuclei becomes efficient.~Rather than reversing the contraction, the energy from the oxygen ignition is consumed by this process, and the star collapses to form a BH.~The \emph{upper edge} of the BHMG corresponds to the stars with the lightest CO core mass that can collapse to a BH via this process.~

The location of this transition is set by the competition between the energy released during oxygen burning and the energy lost to photodisintegration.~The former depends sensitively on the carbon-to-oxygen ratio established during core helium burning \citep{Farmer:2020xne}, while the latter depends exponentially on the core temperature during collapse \citep{Rauscher:2010pu}.~As a result, the upper edge is sensitive both to helium-burning reaction rates (including $^{12}\mathrm{C}(\alpha,\gamma)^{16}\mathrm{O}$), which set the core composition, and to oxygen-burning processes that directly influence the collapse dynamics.

\section{Methods}
We simulate the evolution of isolated non-rotating helium cores from the zero-age horizontal branch (ZAHB) to the endpoint of their evolution, either complete disruption via PISN or collapse to a BH via the photodisintegration instability.~Our simulations are performed using the stellar evolution code MESA (version 12778)~\citep{Paxton:2010ji,Paxton:2013pj,Paxton:2015jva,Paxton:2017eie,Paxton:2019lxx,MESA:2022zpy}.~MESA is a one-dimensional hydrostatic code equipped with an HLLC hydrodynamical solver that enables it to simulate (P)PISN.~We refer the reader to~\citet{Marchant:2018kun,2019ApJ...887...53F,Croon:2020oga,Farmer:2020xne,Farag:2022jcc,Croon:2023kct,Croon:2025gol} for further details.~The code employed here is included in the paper's reproduction package, which will be made available upon publication.

We explore the effects of environmental processes, stellar input physics, and numerics by adopting a fiducial model and varying each parameter individually.~Our fiducial model adopts parameters that are commonly used to describe massive stars undergoing (P)PISN below the lower edge of the BHMG.~Our fiducial model has metallicity $Z=10^{-5}$, and the numerical controls correspond to those needed to well-resolve the lower BHMG edge \citep{Mehta:2021fgz,Farag:2022jcc,Croon:2023kct}.~Other parameter choices are described below.~A summary of processes varied, ranges explored, and fiducial values is given in table~\ref{tab:summary}.~

\begin{table}
\centering
\caption{Stellar processes and numerical controls varied to assess systematic uncertainties.~More detail in text.
}
\begin{tabular}{lll}
\hline
\textbf{Process} & \textbf{Fiducial} & \textbf{Range explored} \\
\hline
\multicolumn{3}{l}{\textit{Nuclear rates}} \\
\hline
$^{12}$C$(\alpha,\gamma)^{16}$O 
    & $\mathbf{0\sigma}$ 
    & $\pm 3\sigma$ \\

$3\alpha$ 
    & $\mathbf{0\sigma}$ 
    & $\pm 3\sigma$ \\

$^{16}$O+$^{16}$O 
    & $\mathbf{1}$ 
    & $0.1\,\text{--}\,10$ \\

$^{12}$C+$^{12}$C 
    & $\mathbf{1}$ 
    & $0.1\,\text{--}\,10$ \\

\hline
\multicolumn{3}{l}{\textit{Stellar physics}} \\
\hline
Metallicity $Z$ 
    & $\mathbf{10^{-5}}$ 
    & $10^{-5}\,\text{--}\,10^{-3}$ \\

Wind efficiency $\eta_{\rm wind}$ 
    & $\mathbf{1.0}$ 
    & $0.4\,\text{--}\,1.0^{\dagger}$ \\

Mixing length $\alpha_{\rm MLT}$ 
    & $\mathbf{2.0}$ 
    & $1.5\,\text{--}\,2.5$ \\

Overshoot $f_{\rm ov}$ 
    & $\mathbf{0.01}$ 
    & $0.001\,\text{--}\,0.05$ \\

Neutrino losses 
    & $\mathbf{0\sigma}$ 
    & $\pm 3\sigma$ \\

$\sin^2\theta_W$ 
    & $\mathbf{0.2319}$ 
    & $0.2223,\,0.23867$ \\

\hline
\multicolumn{3}{l}{\textit{Numerics}} \\
\hline
{\tt delta\_lgRho\_cntr\_limit}
    & $\mathbf{0.001}$ 
    & $0.001\,\text{--}\,0.00025$ \\

{\tt varcontrol\_target} 
    & $\mathbf{5\times10^{-4}}$ 
    & $5\times10^{-4}\,\text{--}\,5\times10^{-5}$ \\

{\tt max\_dq }
    & $\mathbf{5\times10^{-4}}$ 
    & $5\times10^{-4}\,\text{--}\,5\times10^{-6}$ \\

{\tt mesh\_delta\_coeff }
    & $\mathbf{0.8}$ 
    & $0.8\,\text{--}\,0.1$ \\

{\tt split\_merge\_amr\_nz\_baseline}
    & $\mathbf{6000}$ 
    & $6000\,\text{--}\,12000$ \\

{\tt scale\_max\_correction}
    & $\mathbf{1}$ 
    & $1\,\text{--}\,0.05$ \\

\hline
\end{tabular}
\begin{flushleft}
$^{\dagger}$ Wind variations were performed at  metallicity $Z=5\times10^{-4}$ since winds are negligible at the fiducial $Z=10^{-5}$.
\end{flushleft}
\label{tab:summary}
\end{table}

For each parameter configuration, we locate the upper edge of the BHMG by evolving ZAHB models over an initially broad range of helium-core masses and iteratively refining the mass interval that brackets the transition between PISN and core collapse.~We continue this refinement until the transition is resolved with a mass spacing of $1\,\msun$.~We define the upper edge as the BH remnant mass of the lowest initial-mass model that avoids full disruption by a PISN.~The BH mass is taken to be the mass of bound stellar material at core collapse.

\subsection{Nuclear Physics}\label{sec:nuclearphysics}

As noted above, the upper edge is set by the competition between PISN and photodisintegration, with the former being sensitive to the C/O ratio at core helium depletion and the latter being sensitive to the temperature during the collapse.~We explore several reactions that can influence these within the {\tt approx\_21} reaction network.

\subsubsection{$^{12}\mathrm{C}(\alpha,\gamma)^{16}\mathrm{O}$}

The rate of the $^{12}\mathrm{C}(\alpha,\gamma)^{16}\mathrm{O}$ reaction, active during core helium burning, determines the final C/O ratio \citep{Farmer:2020xne}.~Larger rates reduce this ratio, causing more violent PISN explosions via the increased amount of $^{16}$O {and the lower $^{12}$C abundance available to form a convective $^{12}$C burning shell to counteract contraction.~Similarly, smaller rates result in less violent explosions.~

In this work we adopt the state-of-the-art $^{12}\mathrm{C}(\alpha,\gamma)^{16}\mathrm{O}$ rate tables from \cite{deBoer:2017ldl}, which were updated by \cite{Mehta:2021fgz} to have a finer resolution, and varied them over $\pm3\sigma$ around the median $R_{\rm med}(T)$ using $R_\sigma(T)=R_{\rm med}(T)\exp[\sigma\mu(T)]$, where $\mu(T)$ represents the temperature-dependent uncertainty under a log-normal assumption \citep{deBoer:2017ldl,Farmer:2020xne,Mehta:2021fgz}.

\subsubsection{Triple-$\alpha$ process}

The triple-$\alpha$ ($3\alpha$) process sets the lifetime of core helium burning.~A stronger rate shortens this phase, leaving less time for the $^{12}\mathrm{C}(\alpha,\gamma)^{16}\mathrm{O}$ reaction to operate.~This leads to an increase in the C/O ratio at core-helium depletion, which results in weaker PISN explosions.~The converse is true for weaker rates.

Our fiducial model adopts the MESA default rate, which is taken from the {\tt NACRE} library \citep{1999NuPhA.656....3A}.~This does not provide uncertainties, so we varied the rate over $\pm3\sigma$ using the tables from \citet{2019ApJ...887...53F,Farmer:2020xne}, constructed by sampling STARLIB \citep{2013ApJS..207...18S} using the same log-normal prescription as above.

\subsubsection{$^{16}\mathrm{O}{}^{16}\mathrm{O}$}

The $^{16}\mathrm{O}+^{16}\mathrm{O}$ reaction rate affects the location of the upper edge through its impact on photodisintegration.~Stronger oxygen burning releases more energy during the onset of contraction, increasing pressure support and limiting the compressional rise in core temperature.~Since photodisintegration rates scale approximately as $T^{3/2}\exp(-Q/k_BT)$, with $Q$ the relevant threshold energy \citep{Rauscher:2010pu}, this suppresses photodisintegration and shifts the upper edge to higher masses.~The converse holds for weaker oxygen burning.~

The {\tt approx\_21} reaction network approximates steady-state oxygen burning using a single effective $^{16}\mathrm{O}+^{16}\mathrm{O}$ process that combines multiple reaction channels.~As a result, this rate does not correspond to a single nuclear cross section, and its uncertainty is not uniquely specified.~We therefore explore its impact by applying a global scaling factor $f_{16\mathrm{O}} \in [0.1, 10]$, following \citep{2019ApJ...887...53F,Xin:2026zyt}.

\subsubsection{$^{12}\mathrm{C}+^{12}\mathrm{C}$}
{Variations in the $^{12}\mathrm{C}+^{12}\mathrm{C}$ rate primarily affect the core structure during earlier burning stages, and are therefore expected to have a more indirect impact on the upper edge.}~We vary the rate of this compound reaction by applying a scaling factor $f_{12\mathrm{C}} = 0.1, 1, 10$.~

\subsection{Stellar Winds}

Stellar winds influence the location of the upper edge by reducing the final CO core mass, thereby modifying the conditions for pair-instability and photodisintegration at collapse.~Stronger winds are therefore expected to shift the upper edge to lower BH masses.~Because our models begin from ZAHB helium cores, the metallicity dependence studied here primarily reflects mass loss during late stellar evolution.

We adopt the wind prescription of \citet{Brott:2011ni}, which combines the hot-star wind prescription of \citet{1995A&A...299..151H}~---~scaled by a factor of $\langle \rho^2 \rangle/\langle \rho \rangle^2 =0.1$ to account for clumping \citep{2010ApJ...725..940Y}~---~with the cool-star winds of \citet{1990A&A...231..134N}.~The mass-loss rate scales with metallicity as $Z^{0.85}$ \citep{2000A&A...362..295V}.~

\subsubsection{Metallicity}
We vary the metallicity over the range $Z \in [10^{-5}, 10^{-3}]$, spanning Population III to Population II stars.

\subsubsection{Wind efficiency}

We scale the wind mass-loss rate by a factor $\eta \in [0.4, 1.0]$.~Since our fiducial metallicity $Z = 10^{-5}$ leads to negligible wind mass loss, we instead adopt $Z = 5\times10^{-4}$ for this experiment.

\subsection{Mixing}

Mixing of fresh material into the core can prolong core-helium burning and increase the final CO core mass, while also altering the C/O ratio at core helium depletion.~We explore the efficiency of two mixing processes.

\subsubsection{Convective mixing}

We vary the {mixing length parameter}~$\alpha_{\rm MLT}$ in the range $1.5\le\alpha_{\rm MLT}\le 2.5$ with the fiducial value $\alpha_{\rm MLT}= 2.0$.~We adopt the Cox prescription \citep{1968pss..book.....C}.

\subsubsection{Overshooting}

We include exponential diffusive overshooting, in which convective mixing extends beyond the formal convective boundary with a diffusion coefficient that decays exponentially over a characteristic length-scale set by $f_{\rm ov}H_p$ (where $H_p$ is the local pressure scale height).~The transition from convective mixing to overshooting then begins a distance $f_0H_p$ inside the convective boundary.~We adopt $f_{\rm ov}=0.01$ as our fiducial value and vary it over the range $10^{-3}\le f_{\rm ov}\le0.05$, while fixing $f_0=0.005$.

\subsection{Neutrinos}

Neutrino emission removes energy from the stellar core.~This requires stable nuclear burning to compensate, affecting the lifetime of core helium burning.~Increased neutrino losses lead to more rapid cooling of the core, requiring higher burning rates to maintain equilibrium and typically resulting in an increased C/O at core-helium depletion \citep{Heger:2008er,Croon:2020ehi,Croon:2020oga,Sakstein:2020axg,Fiorillo:2026vkd}.

\subsubsection{Neutrino loss rates}

We vary the neutrino loss rate fits of \citet{1996ApJS..102..411I} by $\pm1,\,2,$ and $3\sigma$.~The quoted errors for each process are 10\% (pair emission),
1\% (photo-production), 5\% (plasma),  10\% (recombination).

\subsubsection{Weinberg angle}
{An increase in the Weinberg angle enhances neutrino emission rates through its effect on the weak interaction (vector-) couplings, leading to more efficient cooling.}~We explore three different reported values of the Weinberg angle:~our fiducial value of 0.2319 \citep{1996ApJS..102..411I}, 0.23867 \citep{Erler:2004in}, and 0.2223 \citep{Mohr2016}.~

\subsection{Numerics}
Lastly, we vary several numerical controls described below.~The fiducial values correspond to the minimum needed to well-resolve the lower BHMG edge \citep{Mehta:2021fgz,Farag:2022jcc,Croon:2023kct}, with all variations exploring higher numerical resolution.~These variations allow us to assess whether the inferred upper edge is numerically converged.

\subsubsection{Nuclear Reaction Network}
Our fiducial model adopts the {\tt approx\_21} network, which follows $\alpha$-chain reactions from carbon to iron.~The network compounds several $(\alpha,\gamma)$ and $(p,\gamma)$ reactions into $^{12}$C$^{16}$O, $^{12}$C$^{12}$C, and $^{16}$O$^{16}$O processes for computational efficiency.~We explore the larger {\tt mesa\_75} and {\tt mesa\_128} networks, which have 75 and 128 isotopes up to zinc-60 respectively.

\subsubsection{Timestep}

We vary {\tt delta\_lgRho\_cntr\_limit}, which limits changes in the central density between timesteps, over the range $0.001$--$0.00025$.~Smaller values enforce smaller timesteps.~
We also vary {\tt varcontrol\_target} over the range $5\times10^{-4}$--$5\times10^{-5}$.~This limits the allowed variation in stellar properties between timesteps.~Smaller values limit changes to smaller values.

\subsubsection{Spatial resolution}
We vary {\tt max\_dq} over the range $5\times10^{-4}$--$5\times10^{-6}$.~This control sets the maximum cell size in units of the star's mass.~Smaller values enforce a larger number of cells.

We also vary {\tt mesh\_delta\_coeff} over the range $0.8$--$0.1$.~This control sets the maximum
allowed change in quantities between adjacent
mesh points.~Smaller values give a finer mesh.

\subsubsection{Mesh refinement}

We vary {\tt split\_merge\_amr\_nz\_baseline} between 6000 and 12000.~This control sets the target number of zones when adaptive mesh refinement is active.

\subsubsection{Solver}

We vary {\tt scale\_max\_correction} over the range $1$--$0.05$.~This parameter limits the maximum relative correction applied to stellar variables during each timestep.~Smaller values enforce more conservative updates, improving numerical stability and the robustness of the solution.

\begin{figure*}
    \centering
    \includegraphics[width=0.97\linewidth]{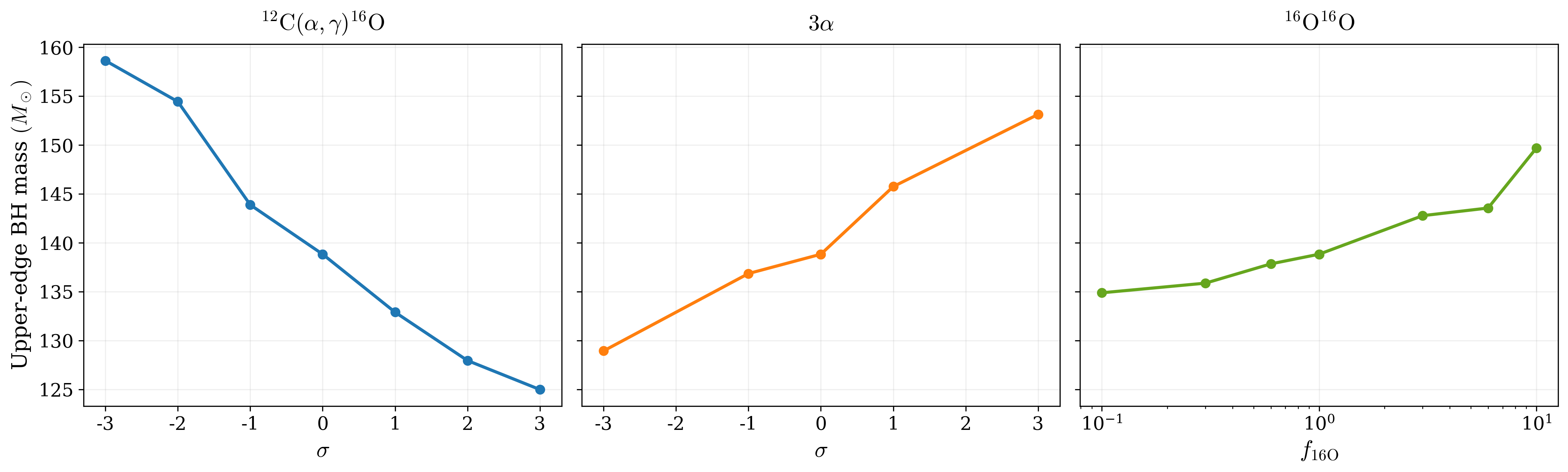}
    \caption{Variation of the upper edge of the BHMG with nuclear rates, as described in text.~
    }
    \label{fig:rates}
\end{figure*}

\section{Results}

We summarize the results of our experiments in Fig.~\ref{fig:biggestshifts} for all processes whose variations yield shifts of $>1\,\msun$ in the location of the upper BHMG edge.~Several notable trends emerge:

\begin{enumerate}
    \item 
    {\bf Nuclear rates dominate the uncertainty.}

As for the lower edge, the location of the upper edge is highly sensitive to variations in nuclear reaction rates active during core helium burning.~The largest effect arises from the $^{12}\mathrm{C}(\alpha,\gamma)^{16}\mathrm{O}$ rate, which produces shifts of $\Delta M \simeq 30\,\msun$, followed by the triple-$\alpha$ process with $\Delta M \simeq 25\,\msun$.
Variations of the $^{16}\mathrm{O}+^{16}\mathrm{O}$ reaction rate yield shifts of $\Delta M \simeq 15\,\msun$;~similar shifts were reported by \cite{Xin:2026zyt}.~We show the upper edge BH mass as a function of rate variation for each of these processes in figure~\ref{fig:rates}.~{The physical origin of these trends is discussed in Section~\ref{sec:nuclearphysics}.}

\vspace{2mm}
\item 
    {\bf The $^{16}\mathrm{O}+^{16}\mathrm{O}$ rate controls the BHMG width.}

The $^{16}\mathrm{O}+^{16}\mathrm{O}$ rate has little effect on the lower BHMG edge, where the outcome is primarily controlled by the onset of pair instability and whether explosive oxygen burning unbinds the star \citep{2019ApJ...887...53F}.~However, it plays a crucial role in determining the peak temperature reached during collapse.~Since photodisintegration rates are exponentially sensitive to temperature, the peak temperature strongly impacts whether collapse is reversed into an explosion or proceeds to BH formation.

Since the $^{16}\mathrm{O}+^{16}\mathrm{O}$ rate shifts the upper edge, while leaving the lower edge largely unchanged, it plays a key role in determining the \emph{width} of the gap.~This suggests that measurements of the gap width may provide a novel probe of this reaction rate.

\vspace{2mm}
\item 
    {\bf Metallicity contributes an 8$\msun$ uncertainty.}

We find that the location of the upper edge depends on metallicity at the $\sim8\,\msun$ level.~Unlike nuclear reaction rates, metallicity does not represent a fundamental theoretical uncertainty, but rather reflects environmental variation across stellar populations.~It can therefore be considered a source of astrophysical scatter in the location of the upper BHMG edge.~We note that this is a larger scatter than found for the lower edge in similar simulations e.g., \citet{Croon:2023kct} find $3\,\msun$.

\vspace{2mm}
\item 
    {\bf Other uncertainties are subdominant.}

Figure~\ref{fig:biggestshifts} shows that stellar processes such as mixing and stellar winds shift the upper edge by $\lesssim 5\,\msun$.~Moreover, increasing the size of the nuclear reaction network shifts the upper edge by $\lesssim3\,\msun$.~Other processes not shown, such as neutrino losses and the $^{12}{\rm C}^{12}{\rm C}$ rate, have sub-$\msun$ effects.

\vspace{2mm}
\item 
    {\bf Upper edge simulations are well-converged.}

We find that the location of the BHMG upper edge is largely insensitive to the numerical controls we varied.~In contrast, the lower edge of the BHMG is more difficult to resolve, with a complex dependence on numerical choices \citep{Mehta:2021fgz,Farag:2022jcc}.
\end{enumerate}

\section{Discussion and conclusions}

In this work we studied the sensitivity of the location of the upper edge of the black hole mass gap to uncertainties in stellar modeling.~Black holes with masses close to this edge are beginning to be observed through gravitational waves emitted by binary mergers, offering the potential to probe stellar evolution.

Our results, described above, suggest several directions for future work.~First, it would be interesting to study the effects of varying individual $^{16}{\rm O}+^{16}{\rm O}$ reactions, since these are compounded into a single rate in our fiducial model.~Second, varying multiple processes simultaneously would enable covariances to be quantified.~Third, constructing a simulation-informed parametric black hole mass function that can be fit to gravitational wave catalogs (e.g., \citealt{Baxter:2021swn,Ulrich:2024nez}) would allow measurements of the upper edge to be translated into constraints on nuclear reaction rates as more data becomes available.~Finally, it would be interesting to quantify the effects of rotation, {which can both shift the upper edge of the BHMG and introduce a correlated spin signature}.

\section*{Acknowledgments}

D.C.~is supported by STFC Grant No.~ST/T001011/1.~J.S.~is supported by NSF Grant No.~2207880.~J.S.~thanks the IPPP for hospitality and acknowledges an IPPP DIVA fellowship to support the visit.~Our simulations were run on the University of Hawai\okina i's high-performance supercomputer KOA.~The technical support and advanced computing resources from University of Hawai\okina i Information Technology Services – Cyberinfrastructure, funded in part by the NSF MRI award No.~1920304, are gratefully acknowledged.~%


\bibliographystyle{mnras_tex_edited}
\bibliography{refs}

\begin{thebibliography}{}
\makeatletter
\relax
\def\mn@urlcharsother{\let\do\@makeother \do\$\do\&\do\#\do\^\do\_\do\%\do\~}
\def\mn@doi{\begingroup\mn@urlcharsother \@ifnextchar [ {\mn@doi@}
  {\mn@doi@[]}}
\def\mn@doi@[#1]#2{\def\@tempa{#1}\ifx\@tempa\@empty \href
  {http://dx.doi.org/#2} {doi:#2}\else \href {http://dx.doi.org/#2} {#1}\fi
  \endgroup}
\def\mn@eprint#1#2{\mn@eprint@#1:#2::\@nil}
\def\mn@eprint@arXiv#1{\href {http://arxiv.org/abs/#1} {{arXiv:#1}}}
\def\mn@eprint@dblp#1{\href {http://dblp.uni-trier.de/rec/bibtex/#1.xml}
  {dblp:#1}}
\def\mn@eprint@#1:#2:#3:#4\@nil{\def\@tempa {#1}\def\@tempb {#2}\def\@tempc
  {#3}\ifx \@tempc \@empty \let \@tempc \@tempb \let \@tempb \@tempa \fi \ifx
  \@tempb \@empty \def\@tempb {arXiv}\fi \@ifundefined
  {mn@eprint@\@tempb}{\@tempb:\@tempc}{\expandafter \expandafter \csname
  mn@eprint@\@tempb\endcsname \expandafter{\@tempc}}}

\bibitem[\protect\citeauthoryear{{Angulo} et~al.,}{{Angulo}
  et~al.}{1999}]{1999NuPhA.656....3A}
{Angulo} C.,  et~al., 1999, \mn@doi [\nphysa] {10.1016/S0375-9474(99)00030-5},
  \href {https://ui.adsabs.harvard.edu/abs/1999NuPhA.656....3A} {656, 3}

\bibitem[\protect\citeauthoryear{Barkat, Rakavy  \& Sack}{Barkat
  et~al.}{1967}]{Barkat:1967zz}
Barkat Z.,  Rakavy G.,   Sack N.,  1967, \mn@doi [Phys. Rev. Lett.]
  {10.1103/PhysRevLett.18.379}, 18, 379

\bibitem[\protect\citeauthoryear{Baxter, Croon, McDermott  \& Sakstein}{Baxter
  et~al.}{2021}]{Baxter:2021swn}
Baxter E.~J.,  Croon D.,  McDermott S.~D.,   Sakstein J.,  2021, \mn@doi
  [Astrophys. J. Lett.] {10.3847/2041-8213/ac11fc}, 916, L16 (\mn@eprint
  {arXiv} {2104.02685})

\bibitem[\protect\citeauthoryear{Belczynski et~al.}{Belczynski
  et~al.}{2016}]{Belczynski:2016jno}
Belczynski K.,  et~al., 2016, \mn@doi [Astron. Astrophys.]
  {10.1051/0004-6361/201628980}, 594, A97 (\mn@eprint {arXiv} {1607.03116})

\bibitem[\protect\citeauthoryear{Brott et~al.,}{Brott
  et~al.}{2011}]{Brott:2011ni}
Brott I.,  et~al., 2011, \mn@doi [Astron. Astrophys.]
  {10.1051/0004-6361/201016113}, 530, A115 (\mn@eprint {arXiv} {1102.0530})

\bibitem[\protect\citeauthoryear{{Cox} \& {Giuli}}{{Cox} \&
  {Giuli}}{1968}]{1968pss..book.....C}
{Cox} J.~P.,  {Giuli} R.~T.,  1968, {Principles of stellar structure}.
Gordon and Breach, New York

\bibitem[\protect\citeauthoryear{Croon \& Sakstein}{Croon \&
  Sakstein}{2025}]{Croon:2023kct}
Croon D.,  Sakstein J.,  2025, \mn@doi [Phys. Rev. D] {10.1103/53lj-hm4d}, 112,
  063053 (\mn@eprint {arXiv} {2312.13459})

\bibitem[\protect\citeauthoryear{Croon, McDermott  \& Sakstein}{Croon
  et~al.}{2020}]{Croon:2020oga}
Croon D.,  McDermott S.~D.,   Sakstein J.,  2020, \mn@doi [Phys. Rev. D]
  {10.1103/PhysRevD.102.115024}, 102, 115024 (\mn@eprint {arXiv} {2007.07889})

\bibitem[\protect\citeauthoryear{Croon, McDermott  \& Sakstein}{Croon
  et~al.}{2021}]{Croon:2020ehi}
Croon D.,  McDermott S.~D.,   Sakstein J.,  2021, \mn@doi [Phys. Dark Univ.]
  {10.1016/j.dark.2021.100801}, 32, 100801 (\mn@eprint {arXiv} {2007.00650})

\bibitem[\protect\citeauthoryear{Croon, Gerosa  \& Sakstein}{Croon
  et~al.}{2026}]{Croon:2025gol}
Croon D.,  Gerosa D.,   Sakstein J.,  2026, \mn@doi [Mon. Not. Roy. Astron.
  Soc.] {10.1093/mnras/stag073}, 546, stag073 (\mn@eprint {arXiv} {2508.10088})

\bibitem[\protect\citeauthoryear{Erler \& Ramsey-Musolf}{Erler \&
  Ramsey-Musolf}{2005}]{Erler:2004in}
Erler J.,  Ramsey-Musolf M.~J.,  2005, \mn@doi [Phys. Rev. D]
  {10.1103/PhysRevD.72.073003}, 72, 073003 (\mn@eprint {arXiv}
  {hep-ph/0409169})

\bibitem[\protect\citeauthoryear{Farag, Renzo, Farmer, Chidester  \&
  Timmes}{Farag et~al.}{2022}]{Farag:2022jcc}
Farag E.,  Renzo M.,  Farmer R.,  Chidester M.~T.,   Timmes F.~X.,  2022,
  \mn@doi [Astrophys. J.] {10.3847/1538-4357/ac8b83}, 937, 112 (\mn@eprint
  {arXiv} {2208.09624})

\bibitem[\protect\citeauthoryear{{Farmer}, {Renzo}, {de Mink}, {Marchant}  \&
  {Justham}}{{Farmer} et~al.}{2019}]{2019ApJ...887...53F}
{Farmer} R.,  {Renzo} M.,  {de Mink} S.~E.,  {Marchant} P.,   {Justham} S.,
  2019, \mn@doi [Astrophys. J.] {10.3847/1538-4357/ab518b}, \href
  {https://ui.adsabs.harvard.edu/abs/2019ApJ...887...53F} {887, 53} (\mn@eprint
  {arXiv} {1910.12874})

\bibitem[\protect\citeauthoryear{Farmer, Renzo, de Mink, Fishbach  \&
  Justham}{Farmer et~al.}{2020}]{Farmer:2020xne}
Farmer R.,  Renzo M.,  de Mink S.,  Fishbach M.,   Justham S.,  2020, \mn@doi
  [Astrophys. J. Lett.] {10.3847/2041-8213/abbadd}, 902, L36 (\mn@eprint
  {arXiv} {2006.06678})

\bibitem[\protect\citeauthoryear{Fiorillo, Lucente, Sakstein, Vitagliano  \&
  Cantiello}{Fiorillo et~al.}{2026}]{Fiorillo:2026vkd}
Fiorillo D. F.~G.,  Lucente G.,  Sakstein J.,  Vitagliano E.,   Cantiello M.,
  2026 (\mn@eprint {arXiv} {2604.02413})

\bibitem[\protect\citeauthoryear{Fishbach \& Holz}{Fishbach \&
  Holz}{2020}]{Fishbach:2020qag}
Fishbach M.,  Holz D.~E.,  2020, \mn@doi [Astrophys. J. Lett.]
  {10.3847/2041-8213/abc827}, 904, L26 (\mn@eprint {arXiv} {2009.05472})

\bibitem[\protect\citeauthoryear{{Fraley}}{{Fraley}}{1968}]{1968Ap&SS...2...96F}
{Fraley} G.~S.,  1968, \mn@doi [Astrophys. Space Sci.] {10.1007/BF00651498},
  \href {https://ui.adsabs.harvard.edu/abs/1968Ap&SS...2...96F} {2, 96}

\bibitem[\protect\citeauthoryear{{Hamann}, {Koesterke}  \&
  {Wessolowski}}{{Hamann} et~al.}{1995}]{1995A&A...299..151H}
{Hamann} W.-R.,  {Koesterke} L.,   {Wessolowski} U.,  1995, \aap, \href
  {https://ui.adsabs.harvard.edu/abs/1995A&A...299..151H} {299, 151}

\bibitem[\protect\citeauthoryear{Heger, Friedland, Giannotti  \&
  Cirigliano}{Heger et~al.}{2009}]{Heger:2008er}
Heger A.,  Friedland A.,  Giannotti M.,   Cirigliano V.,  2009, \mn@doi
  [Astrophys. J.] {10.1088/0004-637X/696/1/608}, 696, 608 (\mn@eprint {arXiv}
  {0809.4703})

\bibitem[\protect\citeauthoryear{{Itoh}, {Hayashi}, {Nishikawa}  \&
  {Kohyama}}{{Itoh} et~al.}{1996}]{1996ApJS..102..411I}
{Itoh} N.,  {Hayashi} H.,  {Nishikawa} A.,   {Kohyama} Y.,  1996, \mn@doi
  [\apjs] {10.1086/192264}, \href
  {https://ui.adsabs.harvard.edu/abs/1996ApJS..102..411I} {102, 411}

\bibitem[\protect\citeauthoryear{Jermyn et~al.}{Jermyn
  et~al.}{2023}]{MESA:2022zpy}
Jermyn A.~S.,  et~al., 2023, \mn@doi [Astrophys. J. Supp. S.]
  {10.3847/1538-4365/acae8d}, 265, 15 (\mn@eprint {arXiv} {2208.03651})

\bibitem[\protect\citeauthoryear{Marchant, Renzo, Farmer, Pappas, Taam, De~Mink
   \& Kalogera}{Marchant et~al.}{2019}]{Marchant:2018kun}
Marchant P.,  Renzo M.,  Farmer R.,  Pappas K.~M.,  Taam R.~E.,  De~Mink S.~E.,
    Kalogera V.,  2019, The Astrophysical Journal, 882, 36

\bibitem[\protect\citeauthoryear{Mehta, Buonanno, Gair, Miller, Farag, deBoer,
  Wiescher  \& Timmes}{Mehta et~al.}{2022}]{Mehta:2021fgz}
Mehta A.~K.,  Buonanno A.,  Gair J.,  Miller M.~C.,  Farag E.,  deBoer R.~J.,
  Wiescher M.,   Timmes F.~X.,  2022, \mn@doi [Astrophys. J.]
  {10.3847/1538-4357/ac3130}, 924, 39 (\mn@eprint {arXiv} {2105.06366})

\bibitem[\protect\citeauthoryear{Mohr, Newell  \& Taylor}{Mohr
  et~al.}{2016}]{Mohr2016}
Mohr P.~J.,  Newell D.~B.,   Taylor B.~N.,  2016, \mn@doi [Reviews of Modern
  Physics] {10.1103/RevModPhys.88.035009}, 88, 035009

\bibitem[\protect\citeauthoryear{{Nieuwenhuijzen} \& {de
  Jager}}{{Nieuwenhuijzen} \& {de Jager}}{1990}]{1990A&A...231..134N}
{Nieuwenhuijzen} H.,  {de Jager} C.,  1990, \aap, \href
  {https://ui.adsabs.harvard.edu/abs/1990A&A...231..134N} {231, 134}

\bibitem[\protect\citeauthoryear{Paxton, Bildsten, Dotter, Herwig, Lesaffre  \&
  Timmes}{Paxton et~al.}{2011}]{Paxton:2010ji}
Paxton B.,  Bildsten L.,  Dotter A.,  Herwig F.,  Lesaffre P.,   Timmes F.,
  2011, \mn@doi [Astrophys. J. Supp. S.] {10.1088/0067-0049/192/1/3}, 192, 3
  (\mn@eprint {arXiv} {1009.1622})

\bibitem[\protect\citeauthoryear{Paxton et~al.}{Paxton
  et~al.}{2013}]{Paxton:2013pj}
Paxton B.,  et~al., 2013, \mn@doi [Astrophys. J. Supp. S.]
  {10.1088/0067-0049/208/1/4}, 208, 4 (\mn@eprint {arXiv} {1301.0319})

\bibitem[\protect\citeauthoryear{Paxton et~al.}{Paxton
  et~al.}{2015}]{Paxton:2015jva}
Paxton B.,  et~al., 2015, \mn@doi [Astrophys. J. Supp. S.]
  {10.1088/0067-0049/220/1/15}, 220, 15 (\mn@eprint {arXiv} {1506.03146})

\bibitem[\protect\citeauthoryear{Paxton et~al.}{Paxton
  et~al.}{2018}]{Paxton:2017eie}
Paxton B.,  et~al., 2018, \mn@doi [Astrophys. J. Supp. S.]
  {10.3847/1538-4365/aaa5a8}, 234, 34 (\mn@eprint {arXiv} {1710.08424})

\bibitem[\protect\citeauthoryear{Paxton et~al.}{Paxton
  et~al.}{2019}]{Paxton:2019lxx}
Paxton B.,  et~al., 2019, \mn@doi [Astrophys. J. Supp. S.]
  {10.3847/1538-4365/ab2241}, 243, 10 (\mn@eprint {arXiv} {1903.01426})

\bibitem[\protect\citeauthoryear{{Rakavy} \& {Shaviv}}{{Rakavy} \&
  {Shaviv}}{1967}]{1967ApJ...148..803R}
{Rakavy} G.,  {Shaviv} G.,  1967, \mn@doi [Astrophys. J.] {10.1086/149204},
  \href {https://ui.adsabs.harvard.edu/abs/1967ApJ...148..803R} {148, 803}

\bibitem[\protect\citeauthoryear{Rauscher}{Rauscher}{2011}]{Rauscher:2010pu}
Rauscher T.,  2011, \mn@doi [Int. J. Mod. Phys. E] {10.1142/S021830131101840X},
  20, 1071 (\mn@eprint {arXiv} {1010.4283})

\bibitem[\protect\citeauthoryear{Sakstein, Croon, McDermott, Straight  \&
  Baxter}{Sakstein et~al.}{2020}]{Sakstein:2020axg}
Sakstein J.,  Croon D.,  McDermott S.~D.,  Straight M.~C.,   Baxter E.~J.,
  2020, \mn@doi [Phys. Rev. Lett.] {10.1103/PhysRevLett.125.261105}, 125,
  261105 (\mn@eprint {arXiv} {2009.01213})

\bibitem[\protect\citeauthoryear{{Sallaska}, {Iliadis}, {Champange}, {Goriely},
  {Starrfield}  \& {Timmes}}{{Sallaska} et~al.}{2013}]{2013ApJS..207...18S}
{Sallaska} A.~L.,  {Iliadis} C.,  {Champange} A.~E.,  {Goriely} S.,
  {Starrfield} S.,   {Timmes} F.~X.,  2013, \mn@doi [\apjs]
  {10.1088/0067-0049/207/1/18}, \href
  {https://ui.adsabs.harvard.edu/abs/2013ApJS..207...18S} {207, 18} (\mn@eprint
  {arXiv} {1304.7811})

\bibitem[\protect\citeauthoryear{Ulrich, Croon, Sakstein  \& McDermott}{Ulrich
  et~al.}{2024}]{Ulrich:2024nez}
Ulrich Y.,  Croon D.,  Sakstein J.,   McDermott S.,  2024 (\mn@eprint {arXiv}
  {2406.06109})

\bibitem[\protect\citeauthoryear{{Vink}, {de Koter}  \& {Lamers}}{{Vink}
  et~al.}{2000}]{2000A&A...362..295V}
{Vink} J.~S.,  {de Koter} A.,   {Lamers} H.~J.~G.~L.~M.,  2000, \mn@doi [\aap]
  {10.48550/arXiv.astro-ph/0008183}, \href
  {https://ui.adsabs.harvard.edu/abs/2000A&A...362..295V} {362, 295}
  (\mn@eprint {arXiv} {astro-ph/0008183})

\bibitem[\protect\citeauthoryear{{Woosley}}{{Woosley}}{2017}]{2017ApJ...836..244W}
{Woosley} S.~E.,  2017, \mn@doi [Astrophys. J.] {10.3847/1538-4357/836/2/244},
  \href {https://ui.adsabs.harvard.edu/abs/2017ApJ...836..244W} {836, 244}
  (\mn@eprint {arXiv} {1608.08939})

\bibitem[\protect\citeauthoryear{{Woosley}}{{Woosley}}{2019}]{2019ApJ...878...49W}
{Woosley} S.~E.,  2019, \mn@doi [\apj] {10.3847/1538-4357/ab1b41}, \href
  {https://ui.adsabs.harvard.edu/abs/2019ApJ...878...49W} {878, 49} (\mn@eprint
  {arXiv} {1901.00215})

\bibitem[\protect\citeauthoryear{{Woosley}, {Heger}  \& {Weaver}}{{Woosley}
  et~al.}{2002}]{2002RvMP...74.1015W}
{Woosley} S.~E.,  {Heger} A.,   {Weaver} T.~A.,  2002, \mn@doi [Reviews of
  Modern Physics] {10.1103/RevModPhys.74.1015}, \href
  {https://ui.adsabs.harvard.edu/abs/2002RvMP...74.1015W} {74, 1015}

\bibitem[\protect\citeauthoryear{Xin, Hou, Zhang, Bi  \& Zhao}{Xin
  et~al.}{2026}]{Xin:2026zyt}
Xin W.,  Hou X.,  Zhang X.,  Bi S.,   Zhao G.,  2026, ]
  {10.1088/1674-4527/ae56dd} (\mn@eprint {arXiv} {2603.19883})

\bibitem[\protect\citeauthoryear{{Yoon}, {Woosley}  \& {Langer}}{{Yoon}
  et~al.}{2010}]{2010ApJ...725..940Y}
{Yoon} S.-C.,  {Woosley} S.~E.,   {Langer} N.,  2010, \mn@doi [\apj]
  {10.1088/0004-637X/725/1/940}, \href
  {https://ui.adsabs.harvard.edu/abs/2010ApJ...725..940Y} {725, 940}
  (\mn@eprint {arXiv} {1004.0843})

\bibitem[\protect\citeauthoryear{deBoer et~al.}{deBoer
  et~al.}{2017}]{deBoer:2017ldl}
deBoer R.~J.,  et~al., 2017, \mn@doi [Rev. Mod. Phys.]
  {10.1103/RevModPhys.89.035007}, 89, 035007 (\mn@eprint {arXiv} {1709.03144})

\makeatother
\end{thebibliography}

\end{document}